\documentclass[conference]{IEEEtran}
\IEEEoverridecommandlockouts
\usepackage{cite}
\usepackage{amsmath,amssymb,amsfonts}
\usepackage{graphicx}
\usepackage{textcomp}
\usepackage{xcolor}

\usepackage{subcaption}
\usepackage{url}
\usepackage{booktabs}
\usepackage{algorithm}
\usepackage{algpseudocode}

\def\BibTeX{{\rm B\kern-.05em{\sc i\kern-.025em b}\kern-.08em
    T\kern-.1667em\lower.7ex\hbox{E}\kern-.125emX}}
\begin{document}

\title{Towards a Self-rescuing System for UAVs \\ Under
GNSS Attack\\
\thanks{
}
}

\author{\IEEEauthorblockN{1\textsuperscript{st} Giulio Rigoni}
\IEEEauthorblockA{\textit{Sapienza, University of Rome} \\
Rome, Italy \\
ORCID: 0000-0003-1901-5758}
\and
\IEEEauthorblockN{2\textsuperscript{nd} Nicola Scremin}
\IEEEauthorblockA{\textit{University of Padua} \\
Padua, Italy \\
nicola.scremin [at] studenti.unipd.it}
\and
\IEEEauthorblockN{3\textsuperscript{rd} Mauro Conti}
\IEEEauthorblockA{\textit{University of Padua} \\
Padua, Italy \\
mauro.conti [at] unipd.it}
}

\maketitle

\begin{abstract}
There has been substantial growth in the UAV market along with an expansion in their applications. However, the successful execution of a UAV mission is very often dependent on the use of a GNSS. Unfortunately, the vulnerability of GNSS signals, due to their lack of encryption and authentication, poses a significant cybersecurity issue. This vulnerability makes various attacks, particularly the ``GNSS spoofing attack," and ``GNSS jamming attack'' easily executable. Generally speaking, during this attack, the drone is manipulated into altering its path, usually resulting in an immediate forced landing or crash. 
As far as we know, we are the first to propose a lightweight-solution that enable a drone to autonomously rescue itself, assuming it is under GNSS attack and the GNSS is no longer available, and return safely to its initial takeoff position, thereby preventing any potential crashes.
During the flight, wind plays a critical role as it can instantaneously alter the drone's position. To solve this problem, we have devised a highly effective 2-phases solution: (i) Forward Phase, for monitoring and recording the forward journey, and (ii) Backward Phase, that generates a backward route, based on the Forward Phase and wind presence.
The final solution ensures strong performance in consistently returning the drone to the original position, even in wind situations, while maintaining a very fast computation time.
\end{abstract}


\section{Introduction} \label{sec:introduction}
The utilization of Unmanned Aerial Vehicles (UAVs), referred to as ``drones," is increasingly prevalent within contemporary society. There is a multitude of applications for them, ranging from photography, agriculture, land mapping, to military operations. 
The significance of ensuring the safety of drones resides in the susceptibility of these aerial vehicles to hacking or tampering through varying means.
UAV security encompasses the measures implemented to safeguard UAVs against unauthorized access, manipulation, and destruction by individuals not authorized to utilize them. This encompasses the provision of physical security and safeguarding against potential cyber-threats. The imperative task at hand is to corroborate the security and privacy of the data procured by UAVs, while concurrently preserving the UAVs' invulnerability from any adversarial attacks that could lead to their potential malfunction.

In essence, it is imperative to ensure that UAVs remain immune to the perils that could potentially jeopardize their operational integrity. It is imperative to establish robust security protocols aimed at safeguarding the subject operations whilst ensuring their safe operation. 

This paper focuses on developing a lightweight, GNSS-independent, and wind-aware navigation algorithm for drones, by addressing the limitations of existing GNSS-based navigation systems, which can be vulnerable to spoofing and jamming attacks. Additionally, GNSS-based systems may not be available in certain environments, such as dense forests.
Therefore the proposed algorithm is motivated by the following key aspects:
\begin{itemize}
    \item Computational Efficiency: drones have limited battery life, so it is important to minimize the computational resources required for navigation. Online methods, which update the drone's position and heading in real-time, can be computationally expensive. Thus, the proposed a lightweight algorithm for drone trajectory computation.
    \item Camera Avoidance: high-resolution cameras can provide valuable data for navigation, but they also require significant computational resources. To conserve battery life, the proposed algorithm avoids the use of a camera for path tracking.
    \item Wind Awareness: wind can cause significant deviations from the intended drone flight path. To ensure more precise and reliable navigation, the proposed algorithm explicitly accounts for the influence of wind.
\end{itemize}
We also make the following assumption about the GNSS:
\begin{itemize}
    \item The drone has some built-in systems for GNSS attacks detection.
    \item The drone trusts all the satellites it connects to, at the beginning of a flight.
    \end{itemize}
This is a necessary assumption for any GNSS-based navigation system, as the drone would have no means of detecting a spoofing attack if it were connected to a malicious entity from the beginning. 

The contribution of this work are summarized as follows.
\begin{itemize}
    \item We develop a lightweight system to track the forward route of a drone.
    \item We developed and tested different fast-computing approaches for the calculation of a backward route, thus making the drone GNSS attack-resilient.
    \item Our proposed solution considers and counters the effect of the wind, key factor in a safe flight of UAVs. 
\end{itemize}

The rest of this work is organized as follows: Section \ref{sec:related_works} reviews the relevant related works, Section \ref{sec:backgroud_preliminaries} explains preliminary studies and contains important information for the next sections. Section \ref{sec:proposedSolution} presents our solution, while Section \ref{sec:evaluation} evaluate it. Finally,  Section \ref{sec:conslusions} offers conclusions and future research directions.

\section{Related Works} \label{sec:related_works}
UAV path recognition is a challenging problem that has been investigated extensively in recent years. For example, in \cite{paper_1}, the authors use a combination of sensors and a single-camera navigation system to recognize the route. The drone path is defined by several waypoints, and landmarks centralized by those waypoints are carefully chosen at street intersections.
Information-rich trajectories can be planned in continuous 3-D space by optimizing initial solutions, as proposed in \cite{paper_2}. This approach is particularly useful when the value of sensor information is unevenly distributed in the target area and unknown a priori. The authors achieve very high accuracy and efficiency with this online algorithm.
GNSS-denied environments present a unique challenge for UAV navigation. \cite{paper_3} proposes a solution based on 2D Simultaneous Localization and Mapping (SLAM) method, which uses a LIDAR sensor to help the UAV navigate in such environments.
All of these solutions perform effectively in practical scenarios, but they require significant computational resources. Both image processing and the implementation of online machine-learning algorithms can be resource-intensive. Instead, our work focuses on developing a lightweight algorithm that allows the drone to conserve battery primarily for returning to its initial point.
Wind estimation is another important area of research for UAV navigation. In \cite{wind_estimator_1}, the authors propose a method for estimating wind parameters in complex scenarios where a traditional anemometer is difficult to use. They use data logged by the autopilot of a quad-rotor drone. The authors of \cite{wind_estimator_2} try to correlate pitch and roll angles to wind direction and speed with good results.
Mathematical methods for wind estimation have also been proposed. For example, in \cite{wind_estimator_3}, the authors build a stochastic wind model to estimate the wind.
For our purpose, it would be too computationally expensive to indirectly compute wind power. Additionally, we want to minimize errors in estimating the strength of the wind as much as possible. Using an analytic method to calculate wind power would likely introduce more errors than using an anemometer-based estimation.

In many situations where GNSSis unavailable or its signal is disrupted, Inertial Navigation Systems (INS) can be used to enhance navigation systems. 
Authors in \cite{Gyroscope_4}  focus on position and orientation estimation using inertial sensors, combined with other sensors and models (e.g., optimization-based smoothing, filtering and extended Kalman filter) to increase accuracy. 
In \cite{velocityIMU}, the authors develop a robust system for accurately estimating vehicle speed based on Inertial Measurement Unit (IMU) data. By inputting raw IMU data, the system can estimate the vehicle's speed in kilometers per hour (km/h).
IMUs can also be used to improve maritime navigation. For example, in \cite{shipIMU}, the authors conduct a real-world ship experiment utilizing IMU technology to estimate position accurately when GNSSbecomes unavailable. 
Still, those works don't account for wind effects during navigation, nor try to achieve self-rescuing solutions to recover the drone in case of attacks or GNSSmalfunctions, as we do.

Regrettably, based on our understanding, prior studies predominantly employ heavy computational strategies or rely on GNSS-based methods. There has been limited exploration considering factors such as battery capacity and wind presence in scenarios where GNSSsignals are compromised. In response to these gaps, our system is designed to address these specific requirements. In this work, we introduce a lightweight self-rescuing system for UAVs in GNSS-denied environments.

\section{Background and Preliminaries}\label{sec:backgroud_preliminaries}
First of all, an essential aspect of this research is the emphasis on energy consumption. However, calculating this value directly poses significant challenges as shown in \cite{9600034,sorbelli2023wind}. Even the battery indicator displayed by the device offers only an estimate. Thus, estimating energy consumption during a simulation lacks precision and significance, even if calculated in subsequent flights. However, we can roughly gauge and compare the battery consumption by considering algorithms complexity, number of algorithms and in general the resources required for the computation.
Following, Section \ref{sec:Hardware} details all the requisite hardware and setup employed in this study. Following that, Section \ref{sec:logs} elucidates the storage of data within drones and highlights distinctions among various log files. Section \ref{sec:featuresForPathTracking} explains our specific choices of features for UAV path tracking. Lastly, Section \ref{sec:outdoor} documents the experiment conducted for creating our outdoor dataset.

\subsection{Hardware \& Set-up} \label{sec:Hardware}

For this work, we adopted the DJI Mavic 3 Classic due to the capability of accommodating a substantial amount of payloads needed for the real test and data collection. However, it's worth noting that most DJI drones have similar limitations and copyright protections. Finally, DJI is widely considered the largest drone company, with online resources citing a market share of over 70\%.

\begin{figure}
  \begin{subfigure}[t]{0.23\textwidth} 
    \centering
    \includegraphics[width=1.\linewidth]{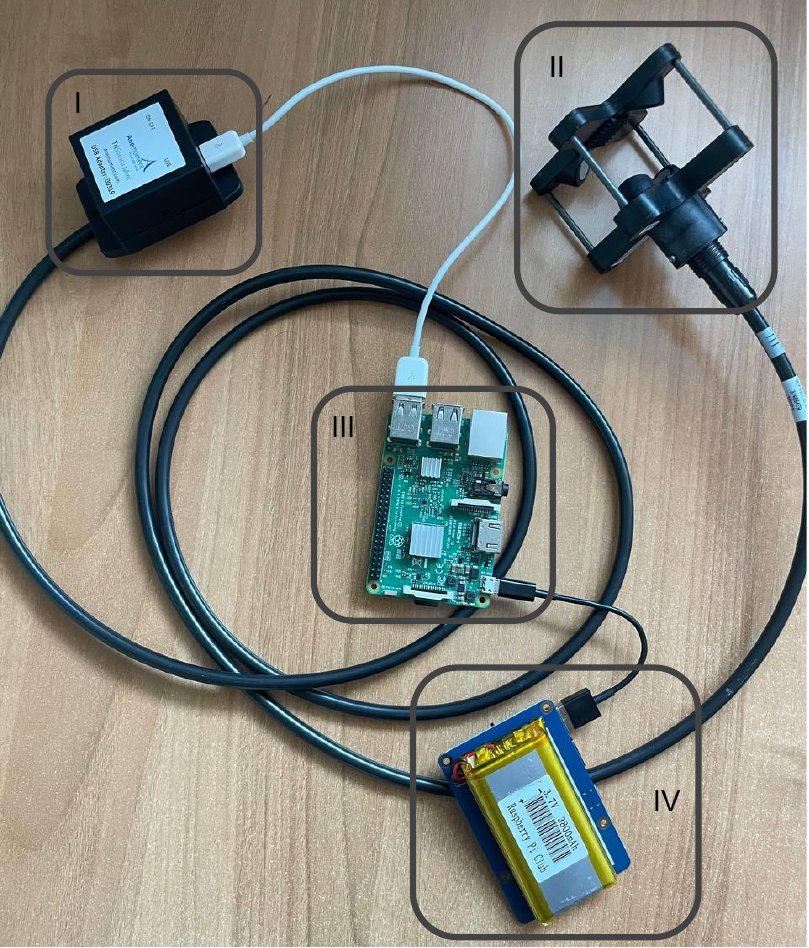} 
    \caption{The external set up mounted on the drone.}
    \label{fig:DroneMountedSetup}
  \end{subfigure}%
  \hfill
  \begin{subfigure}[t]{0.23\textwidth} 
    \centering
    \includegraphics[width=0.985\linewidth]{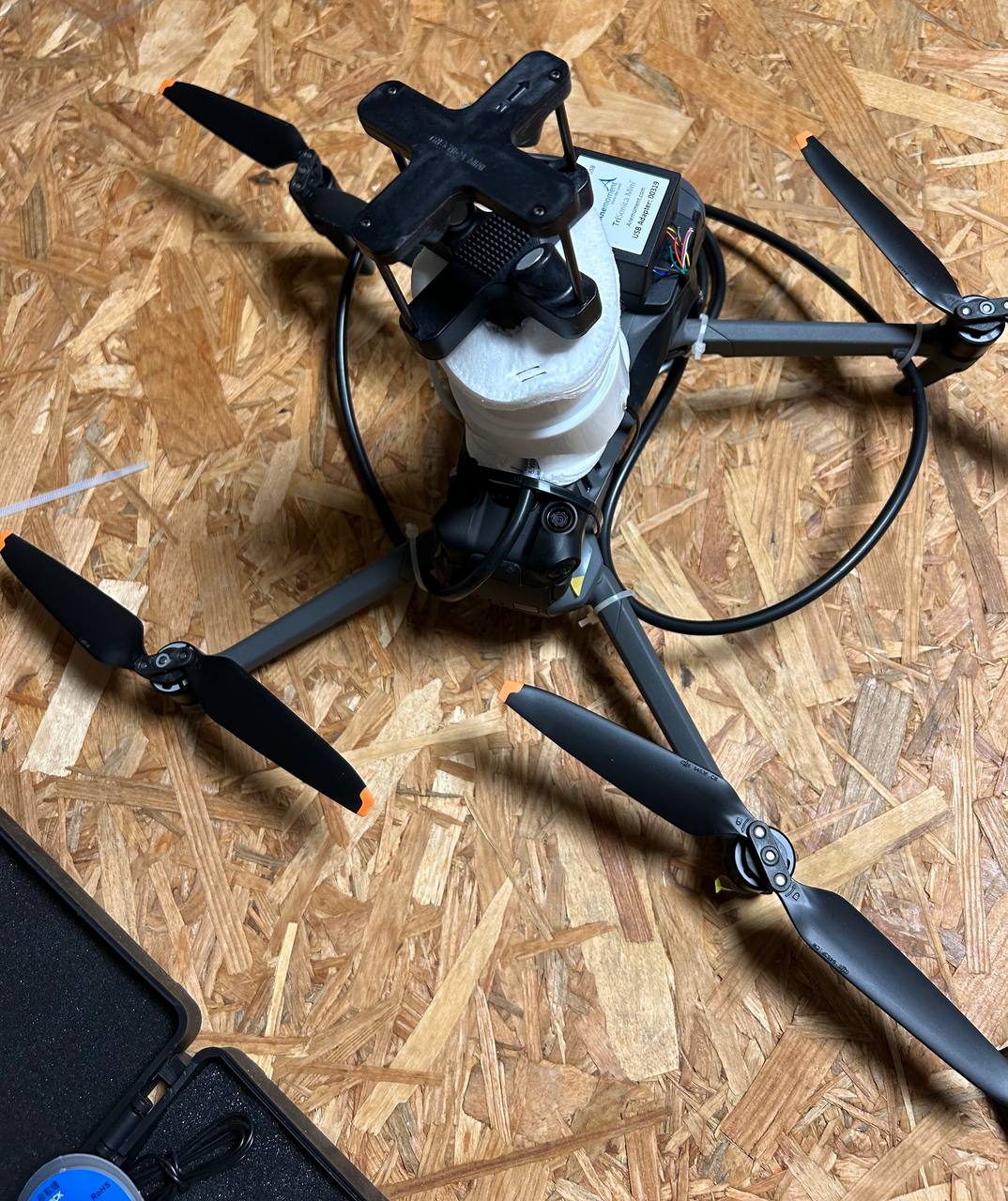} 
    \caption{The Mavic 3 Classic mounted with all the set up.}
    \label{fig:mav3_mounted}
  \end{subfigure}
  \caption{Whole set up utilized: (I) is the USB Adapter, (II) is the Trisonica Anemometer,  (III) shows the Raspberry PI 3 and, (IV) is the power for Raspberry.}
  \label{fig:drone_setup}
\end{figure}

Other than the drone, the full setup includes the following components shown in Figure \ref{fig:DroneMountedSetup}: (i) The TriSonica Mini Wind and Weather Sensor anemometer, which plays a pivotal role by providing wind power measurements. It is a compact anemometer with a measurement path of just 35 mm, weighing less than 50 grams. Despite its small size, it serves as a powerful and highly accurate mobile sensing system designed for atmospheric monitoring, weather reporting, and ecosystem research. (ii) the Raspberry Pi 3, which functions as the system's central processing unit, is responsible for reading and storing all data generated by the Anemometer. It operates as a service that initiates when power is turned on and continues to record data until it is powered off. (iii) A Powerbank, which supplies power to the entire setup throughout the flight. (iv) Lastly, the USB Adapter which establishes a seamless connection between the Raspberry Pi 3 and the Anemometer, ensures smooth communication between the two. Figure \ref{fig:mav3_mounted} depics the full configuration utilized once mounted on the drone.
For the computations and simulations described later in this paper, we utilized an old and computationally slow Acer Aspire 3 laptop ( equipped with an AMD Ryzen 3500U CPU, Radeon Mobile Vega 8 GPU, and 8GB of RAM) in order to ``simulate" the computational power of the adopted drone. The software implementation relied on Python libraries.



\subsection{DJI Log Files}  \label{sec:logs}
DJI Mavic drones produce three types of encrypted logs:

\begin{itemize}

    \item Aircraft .DAT logs contain the most comprehensive and useful data. 
    However, they are not available to the public.

    \item Mobile device .DAT logs are considered a subset of the aircraft .DAT files. 
    and DJI does not provide any support or information about these files, therefore not usable.

    \item Mobile device .TXT logs  are recorded from motor start to motor stop (or signal loss) and include telemetry, battery, gimbal and camera data, flight status, and error flags. However, they do not include motor data. 
    They were initially unreadable, but applications such as Airdata.com\footnote{\url{https://airdata.com/}}, PhantomHelp.com\footnote{\url{https://www.phantomhelp.com/logviewer/upload/}}, and FlightReader\footnote{\url{https://www.flightreader.com/}} can decrypt them and transform them into readable CSV files.
    For this work, we choose to use those log files.
\end{itemize}

\subsection{Features for Path Tracking} \label{sec:featuresForPathTracking}

Driven by the goal of minimizing battery usage for the computation, we sought to utilize readily available information within the drone itself. This led us to analyze the .TXT log file (generated automatically, without incurring additional computational costs, a key benefit of this approach) and the anemometer log, where we discovered critical features contributing to our research aims as described in Section \ref{sec:introduction}. The features we discovered are grouped as follows:

\paragraph{\textbf{Drones Direction Features}}

Inside the drone's log file, we were able to identify three features of interest: \emph{xSpeed, ySpeed} and \emph{height}. The first two express the speed of the drone in miles per hour (MPH), while the last one refers to the flight altitude.
More precisely, xSpeed indicates the speed along the True North, while ySpeed indicates the speed along the True East. In the DJI Developer documentation, it is stated that xSpeed and ySpeed are the current speed of the aircraft in the x/y direction, using the N-E-D (North-East-Down) coordinate system. However, this does not mean that xSpeed is the speed on the x-axis. Rather, it is the speed along the north(+)/south(-) earth frame of reference.
From a closer analysis, we can state that at least the following components are used for x/ySpeed internal calculation, and bear in mind that this information is not disclosed by the manufacturer:

\begin{enumerate}
    \item Compass: A device that consists of a magnetized needle that can pivot to align itself with magnetic north.
    \item Gyroscope: It is a device used for measuring or maintaining orientation and angular velocity \cite{Gyroscope_2,Gyroscope_4}. 
 
    \item Accelerometer: It is a tool that measures proper acceleration and used to determine UAV position \cite{Accelerometer_1, Accelerometer_2}. 
\end{enumerate}

Considering those three components, like previous works in literature \cite{Gyroscope_4,TrajectoryBasedCompassGyrosco,RealTimeVelEstim}, we choose to use the compass to decompose the direction into an “x” and “y” orientation, and the combination of gyroscope and accelerometer to determine the velocity of the drone.

\paragraph{\textbf{Wind Features}} \label{sec:windFeatures}
Inside the anemometer log file,  we found two vectors: V and U; those vectors assess the wind's intensity along the North and East directions concerning the anemometer's position. It's important to note that this does not correspond to the true North and East as per Earth's geographical coordinates. 
The anemometer relies on these vectors, U and V, to compute the wind's magnitude, denoted as $P_w$, using the following formula:
\begin{math}
    P_w = \sqrt{U^2 + V^2}.
\end{math}

Up to this point, we have determined the wind's strength but not its direction. To address this, the anemometer features a compass integrated into the device, but unfortunately, the integrated compass exhibits limited robustness. It's susceptible to interference from the strong magnetic field generated by the drone's propellers, a concern known and raised by the anemometer's manufacturer.
Thus, we positioned the anemometer in such a way that the north “N” aligns with the drone’s forward direction. By construction, it is possible to affirm that the output of the compass and the yaw angle computed by DJI are the same.
However, it's important to acknowledge that the pairs \textit{[xSpeed, ySpeed]} and \textit{[V, U]} pertain to two distinct reference axes. Therefore, it becomes necessary to convert one coordinate system into the other and 
we chose to transform the North/East axis associated with the anemometer into the True North/East reference. The north axis of the anemometer aligns precisely with the yaw angle, denoted as $\gamma$, by design. The east direction consistently maintains a 90° offset from the north, which we'll refer to as $\alpha$. On the true North and East axes, each vector ``V" and ``U" has its components of them. We'll call them ``northV" to represent the component of V aligned with the true north direction and ``eastV" to signify the part of V influencing the true east direction. The same nomenclature applies to ``northU" and ``eastU."

Let's consider as an example the yaw in the range [0, 180], but it is the same for the range [-180, 0):
{\small
\begin{align} \label{eq:start}
    \alpha= \gamma - 90. \\
    northV = cos(\gamma) * V&,\quad eastV = sin(\gamma) * V,\\
    northU = -cos(\alpha) * U&,\quad eastU = -sin(\alpha) * U. 
\end{align}
}

As both of these components have contributions related to True North and True East, we straightforwardly combine them to obtain the final vectors, denoted as $northR$ and $eastR$.:
{\small
\begin{align}\label{eq:toNorthEast}
    northR &= northV + northU, \\
    eastR &= eastV + eastU.
\end{align}}
With the U (eastR) and V (northR) vectors now aligned with the xSpeed and ySpeed directions, the statement ``northR = 10m/s" indicates the presence of a wind component blowing from the north to the south, with a velocity of 10 meters per second.

\subsection{Outdoor Dataset Collection} \label{sec:outdoor}

In the outdoor scenario, we employed the Mavic 3 Classic full setup (see Figure \ref{fig:mav3_mounted}) and we conducted a total of 10 complete flights. All of them were confined to a 300-meter radius, aligning with Italian regulations governing drone ''line-of-sight flights''. Beyond this distance, it would enter a state of ``beyond-line-of-sight" which is prohibited by Italian drone laws.

\section{Proposed Solution}\label{sec:proposedSolution}

In this Section, we will present our solution based on the features described in Section \ref{sec:featuresForPathTracking}. Again, firstly  emphasizing the importance of wind in this problem. Wind is one of the major external factors that can significantly deviate a drone's trajectory. If a drone can learn to compensate for the effects of wind, it can make the return journey much easier by following the same route it took on the outward journey. Thus, our solution consists of two main parts: 

\begin{enumerate}
    \item An algorithm for tracking the forward route autonomously, without relying on GNSS data. This algorithm is developed 
    and tested in both in indoor and outdoor scenario.
    
    \item An algorithm for ensuring the safe return of the drone to its initial location, based on the principles of the forward algorithm. This algorithm is essentially a reversal of the forward algorithm, but it also takes into account the effects of wind.
\end{enumerate}

The combination of both algorithms plus the wind impact calculation, represents the GNSS-independent solution for drone navigation and the explanation of both phases, the Forward Phase in Section \ref{sec:farward} and the backward Phase in Section \ref{sec:backward}, follows.


\subsection{Forward Phase}\label{sec:farward}
\begin{figure}
  \begin{subfigure}{0.24\textwidth} 
    \centering
    \includegraphics[width=\linewidth]{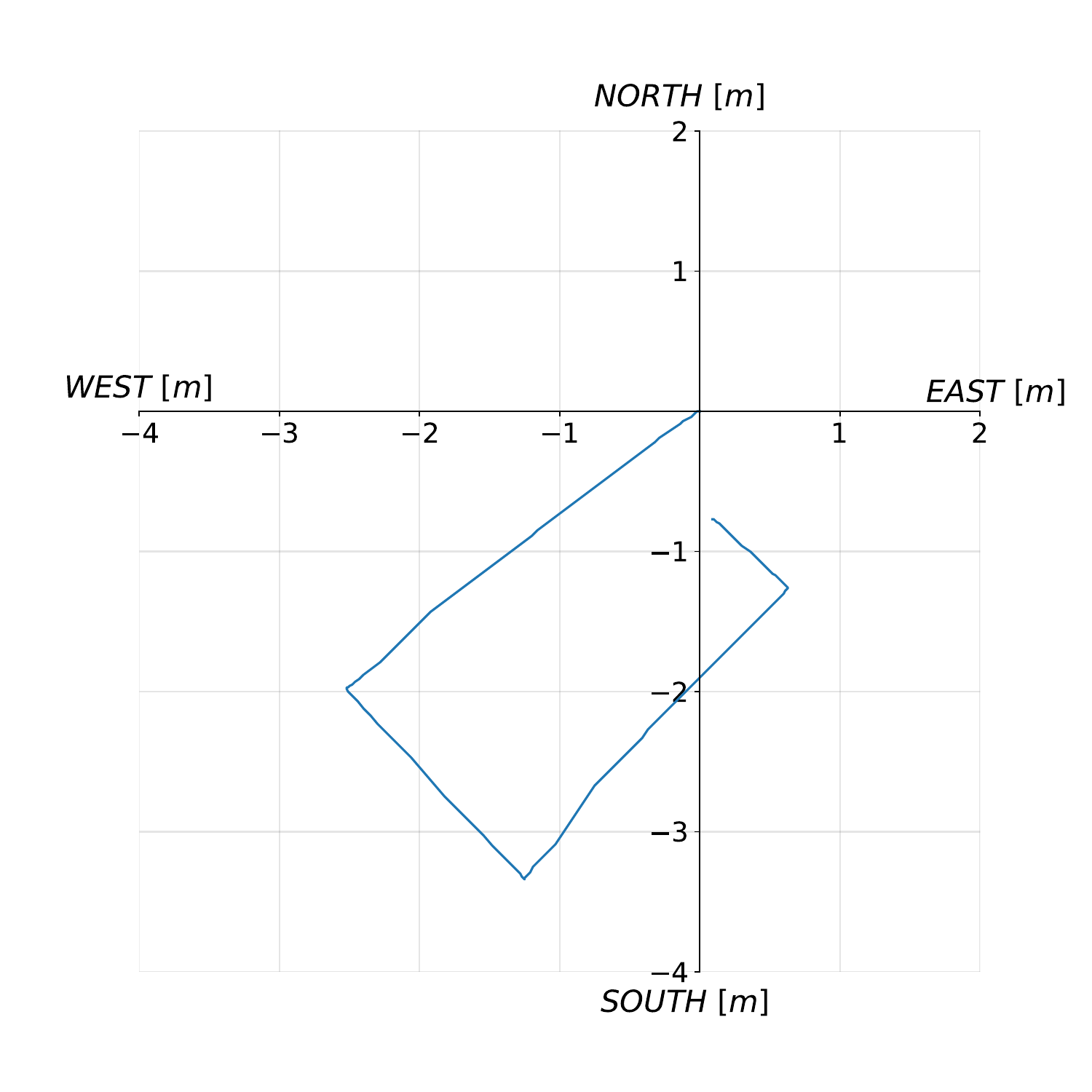} 
    \caption{2D route plotted.}
    \label{fig:2D_plot}
  \end{subfigure}%
  \begin{subfigure}{0.24\textwidth} 
    \centering
    \includegraphics[width=1\linewidth]{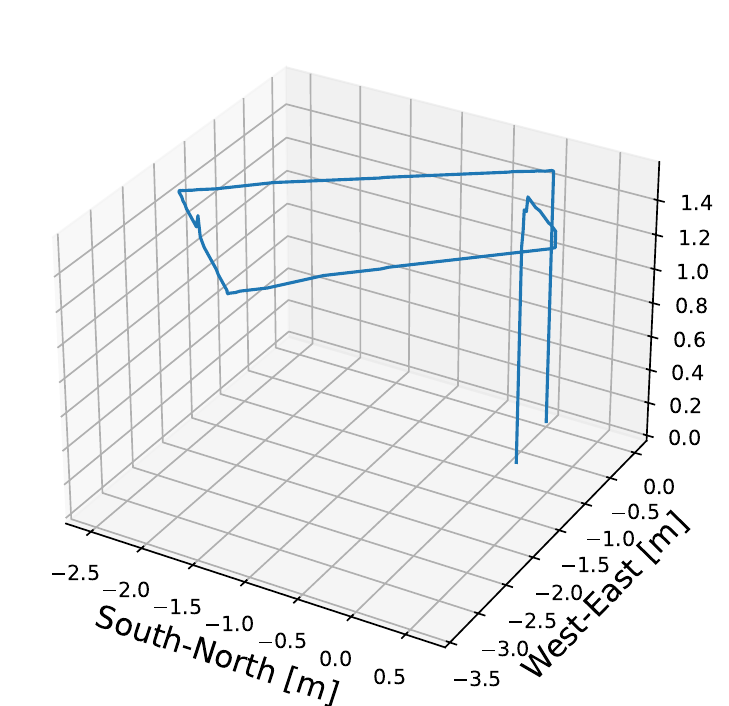} 
    \caption{3D route plotted.}
    \label{fig:3d_route}
  \end{subfigure}
  \caption{It is shown the same route of a flight in a 2D and a 3D space.}
  \label{fig:Rettangolo}
\end{figure}

\begin{figure}[t]
\centering
  \begin{subfigure}[t]{0.23\textwidth} 
    \centering
    \includegraphics[width=0.88\linewidth]{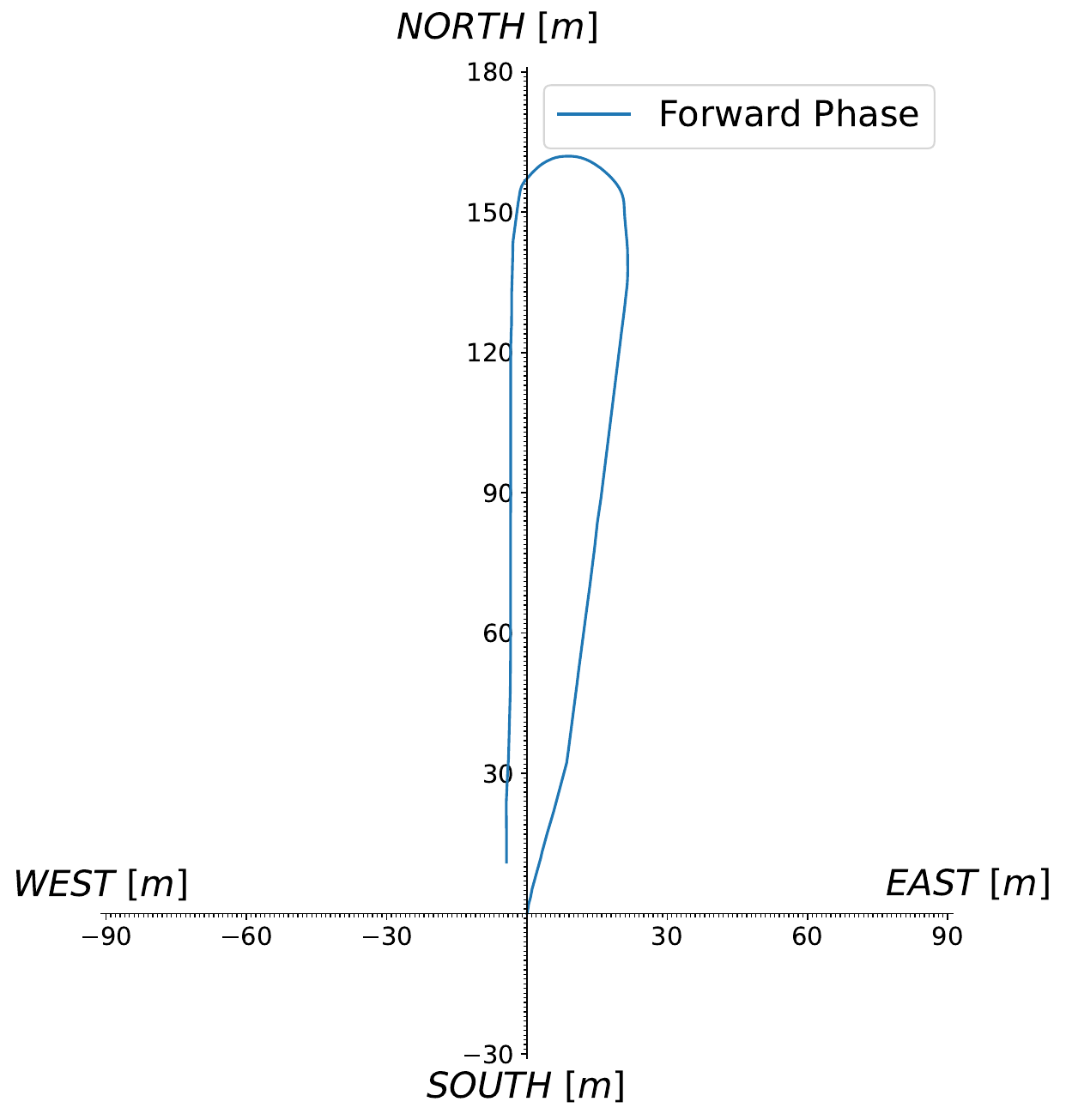} 
    \caption{2D route using xSpeed and ySpeed.}
    \label{fig:proofSpeedReal-a}
  \end{subfigure}%
  \hfill
  \begin{subfigure}[t]{0.23\textwidth} 
    \centering
    \includegraphics[width=\linewidth]{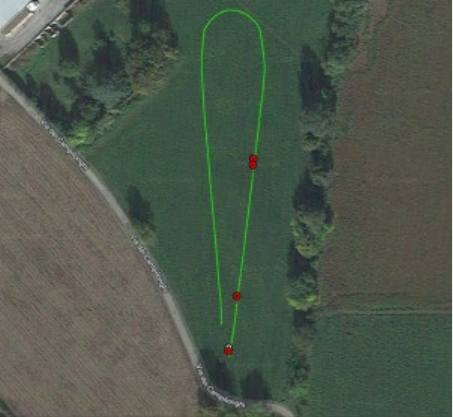}
    \caption{2D route computed by FlightReader Application.}
    \label{}
  \end{subfigure}
  \caption{The same flight but displayed with two different methods.}
  
  \label{fig:proofSpeedReal}
\end{figure}

It is important to stress, that there is no additional computational load imposed on the system during this phase, and a very accurate example of how xSpeed and ySpeed can track the route of the drone in an indoor scenario is shown in Figure \ref{fig:Rettangolo}, while it is also re-proposed for the outdoor scenario in Figure \ref{fig:proofSpeedReal}.
As it is very clear, both figures show how precise is the proposed solution x/ySpeed-based, in reproducing the exact path of the drone.
This could be the answer to our problem, but unfortunately, it is not because we need to consider also the wind blowing in the area and during the flight, therefore the second part of the proposed solution (i.e., the Backward Phase) must factor in the wind impact.

\subsection{Backward Phase}\label{sec:backward}
Before explaining in detail the Backward Phase, let's assume for instance the scenario where there is a total absence of wind: the Backward Phase solution is straightforward, that is simply employ the Forward Phase xSpeed and ySpeed, in reverse order.

Unfortunately, in a real case scenario, the wind is a key factor, thus we need to properly consider its impact on the drone's path. We tried 2 different methods, namely the Weighted Proportion Method and the ML-Based Method, described as follows.

\paragraph{\textbf{Weighted Proportion Method}}  

In our experiments, we observed that the drone autonomously adjusts its speed, at least in part, when encountering wind gusts in order to counter it and stay on track. We cannot pinpoint the precise reasons due to DJI rights restrictions, but based on this assumption, we began by considering that the x/ySpeed already incorporates wind effects to some extent, prompting us to experiment with a proportional approach:
{\small
\begin{align}\label{eq:speed}
    speedF : windF = speedB : windB,
\end{align}
}
where   $speedF$ and $windF$ are the speed and wind power in the forward direction, while $speedB$ and $windB$ correspond to the same parameters but in the context of the return
journey. We can interpret $speedF$ as xSpeed in the forward direction
and $windF$ as the wind acting along the True North direction, also in the forward direction.
Same for ySpeed. Therefore to estimate xSpeed of the return journey starting from eq. \ref{eq:speed} we
have to compute:
{\small
\begin{align}\label{eq:xSpeed_naive}
    xSpeedB = + \frac{xSpeedF * northRB}{northRF},
\end{align}
}
where $northRF$ is the value retrieved using eq. \ref{eq:toNorthEast} in the Forward Phase and $xSpeedF$ is the value computed by the DJI at the same point. Instead, $northRB$ is the value of the anemometer along the true north at the exactly same point during the Forward Phase. All of these values are used to compute the estimated xSpeed, (i.e., $xSpeedB$) that the drone has to take to flight the route in reverse order. 
In eq. \ref{eq:xSpeed_naive}, we are considering equally the propellers' power and the wind speed, but in reality, it is not. In fact, xSpeed and ySpeed are primarily driven by rotor power, with wind speed playing a secondary role. Hence, we have developed a weighted expression in which we assign higher priority to the speed at which the aircraft was traveling during the Forward Phase compared to the intensity
of the wind:
{\small
\begin{align}\label{eq:Completa}
    xSpeedB &= \alpha * xSpeedF + \beta * \frac{xSpeedF * northRB}{northRF} \nonumber. \\
    &= xSpeedF * (\alpha - \beta * \frac{northRB}{northRF}),
\end{align}
}
with $\alpha$ and $\beta$ representing weights. Of course, if  $\beta$=0 means it is not a real case scenario, but the one with total wind absence and where the obtained results are more precise.  However, since we think that DJI builds its own algorithm to contrast the wind, we aim to prioritize the xSpeed and ySpeed values over the wind power measured by the anemometer, as previously stated, setting $\beta$ = 10\% as per our experiments.
Algorithm \ref{alg:backpropagation} is the implementation for the Weighted Proportion Method.

\begin{algorithm}
\caption{Weighted Proportion Method - Backward Phase algorithm using generated data in the Forward Phase to navigate the drone back to the starting point.}\label{alg:backpropagation}
\begin{algorithmic}[1]
\tiny
\Procedure{backPropagation}{velXret, velYret, heightRet, v\_forw, u\_forw, yaw\_forw}
\State \texttt{\% reverse all of these list to compute the backward route}
\State \texttt{\% all of these lists are values stored during the forwarding route}
\For{\texttt{velx, vely, h, v, u, yaw in (velXret, velYret, heightRet, v\_forw, u\_forw, yaw\_forw)}}
    \State \texttt{v\_back $\gets$ readFromAnemometer}
    \State \texttt{\% v\_back $:$ v value retrieved by anemometer on backward route}
    \State \texttt{u\_back $\gets$ readFromAnemometer}
    \State \texttt{\% u\_back $:$ u value retrieved by anemometer on backward route}
    \State \texttt{yaw\_back $\gets$ readFromAnemometer} 
    \State \texttt{\% yaw\_back $:$ yaw angle retrieved by anemometer on backward route}
    \State \texttt{northRF, eastRF $\gets$ convertToNorthEast(v, u, yaw)}
    \State \texttt{\% northRF $:$ wind power along the TRUE north on forwarding route}
    \State \texttt{\% eastRF $:$ wind power along the TRUE east on forwarding route}
    \State \texttt{northRB, eastRB $\gets$ convertToNorthEast(v\_back, u\_back, yaw\_back)}
    \State \texttt{\% northRB $:$ wind power along the TRUE north on backward route}
    \State \texttt{\% eastRB $:$ wind power along the TRUE east on backward route}
    \If \texttt{ northRF != 0:}
    \State \texttt{xSpeed $\gets$ velx ($\alpha$ + $\beta$(northRB/northRF))}
    \Else
    \State \texttt{xSpeed $\gets$ velx}
    \EndIf
    \If \texttt{ eastRF != 0:}
    \State \texttt{ySpeed $\gets$ vely($\alpha$ + $\beta$(eastRB/eastRF))}
    \Else
    \State \texttt{ySpeed $\gets$ vely}
    \EndIf  
\EndFor
\EndProcedure
\end{algorithmic}
\end{algorithm}

\paragraph{\textbf{ML-Based Method}} To potentially improve the results obtained with the Weighted Proportion Method (presented in Section \ref{sec:backwardPhaseEval}), we explore a Machine Learning approach. The idea is to establish the relation between the x/ySpeed obtained from DJI and the NorthR and EastR vectors, defined in Section \ref{sec:windFeatures} and referring to the wind speed directional components. We start by checking the Pearson Correlation Coefficient\footnote{\url{https://en.wikipedia.org/wiki/Pearson_correlation_coefficient62}} between the aforementioned variables. 
Through this approach, we determined that the correlation between xSpeed and northR is approximately $-0.77$, while the correlation between ySpeed and eastR is around $-0.59$. These correlation values indicate a notable relationship between these variables.

\begin{figure}
 \begin{subfigure}{0.23\textwidth}
     \includegraphics[width=\textwidth]{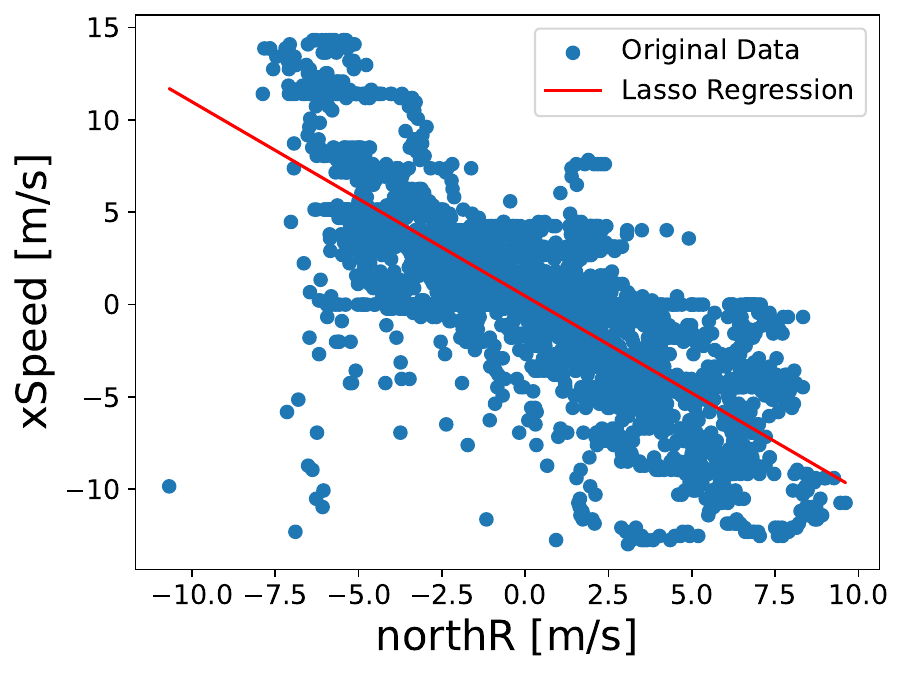}
     \caption{Correlation between northR and xSpeed. In this case, LASSO $R^2$ score is 0.59.}
     \label{fig:lassoNorth}
 \end{subfigure}
 \hfill
 \begin{subfigure}{0.23\textwidth}
     \includegraphics[width=\textwidth]{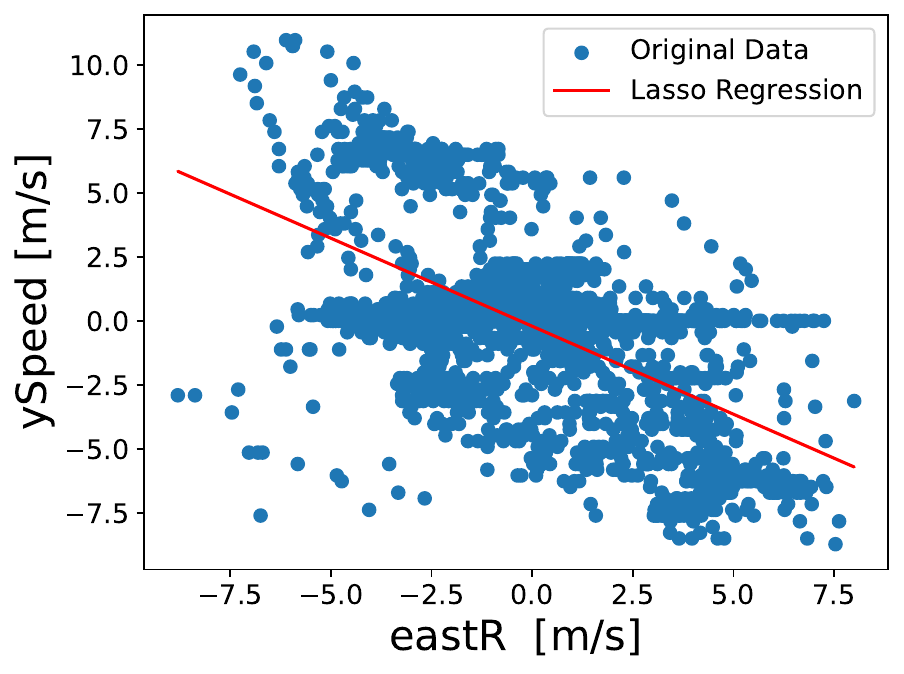}
     \caption{Correlation between northE and ySpeed. In this case, LASSO $R^2$ score is 0.32.}
     \label{fig:lassoEast}
 \end{subfigure}
 \caption{LASSO regression plot for both cases.}
 \label{fig:lasso}
\end{figure}

Due to these observed relationships, we chose 
a regression analysis, as it is closely linked to correlation. We explored various regression methods, including linear, polynomial, exponential, and neural network regressions. Figure \ref{fig:lasso} illustrates the best-performing model, which is the LASSO regression.
In literature, the coefficient of determination R2, can be interpreted similarly to the Pearson Correlation Coefficient, but for LASSO.
In our specific case, we obtained values of 0.59 and 0.32 even considering the presence of outliers visible in Figure \ref{fig:lassoEast}. Notice that in literature, between $0.3$ and $0.5$ is considerate ``moderate influence''. 
Algorithm \ref{alg:backpropagationLasso} implements the LASSO model already trained with all the flights' data. 

\begin{algorithm}[t]
\caption{ML-Based Method - Backward Phase algorithm using a trained Lasso regression model. Given as input the data computed by the anemometer, the LASSO model estimates both the xSpeed and the ySpeed to reach the starting point.}\label{alg:backpropagationLasso}
\begin{algorithmic}[1]
\tiny
\Procedure{backPropagation}{}
\State \texttt{lasso\_north, lasso\_east $\gets$ buildModels()}
\State \texttt{\% lasso\_north $:$ model trained with pair of values (speed, wind) in order to predict xSpeed, given the wind power along the TRUE north}
\State \texttt{\% lasso\_east $:$ model trained with pair of values (speed, wind) in order to predict ySpeed, given the wind power along the TRUE east}
\For{\texttt{v, u, yaw in backwardRoute}}
    \State \texttt{\% v, u and yaw $:$ values retrieved by anemometer during the backward route}
    \State \texttt{northRB, eastRB $\gets$ convertToNorthEast(v, u, yaw)}
    \State \texttt{xSpeed\_pred $\gets$ lasso\_north.predict(northRB)}
    \State \texttt{ySpeed\_pred $\gets$ lasso\_east.predict(eastRB)}
    \State \texttt{\% use the following speeds up to new reading of sensor (for case of the thesis 0.2 second) \%}
\EndFor
\EndProcedure
\end{algorithmic}
\end{algorithm}

\section{Evaluation} \label{sec:evaluation}
Following, we will show the results obtained using the proposed solution explained in Section \ref{sec:proposedSolution}, starting from the Forward Phase in Section \ref{sec:forwardPhaseEval}, and then focusing more on the Backward Phase in Section \ref{sec:backwardPhaseEval}.

\subsection{Forward Phase Evaluation}\label{sec:forwardPhaseEval}
The evaluation of the Forward Phase is straightforward, our solution based on  x/ySpeed is able to keep track and reproduce perfectly the route of the drone during this phase, as shown in Figure \ref{fig:Rettangolo} and Figure \ref{fig:proofSpeedReal}. Both the indoor and outdoor scenario test flights are reproduced perfectly compared to the original path taken by the drone. Important to notice, again, here the wind is not considered yet, because during this phase the drone is still under the pilot's control.

\subsection{Backward Phase Evaluation} \label{sec:backwardPhaseEval}
This Section presents the assessment of the Backward Phase, first examining the Weighted Proportion Method and then the ML-Based Method, as detailed in Section \ref{sec:backward}, notice that in this stage of evaluation, the significance of wind becomes prominent. 
Moreover, given the difficulty and problematic nature of energy consumption computation, as explained in Section \ref{sec:backgroud_preliminaries}, our focus shifts towards the quantity and speed of computation necessary for the proposed solution's usability.

\paragraph{\textbf{Weighted Proportion Method}}
\begin{figure}[]
 \begin{subfigure}{0.24\textwidth}
     \includegraphics[width=\textwidth]{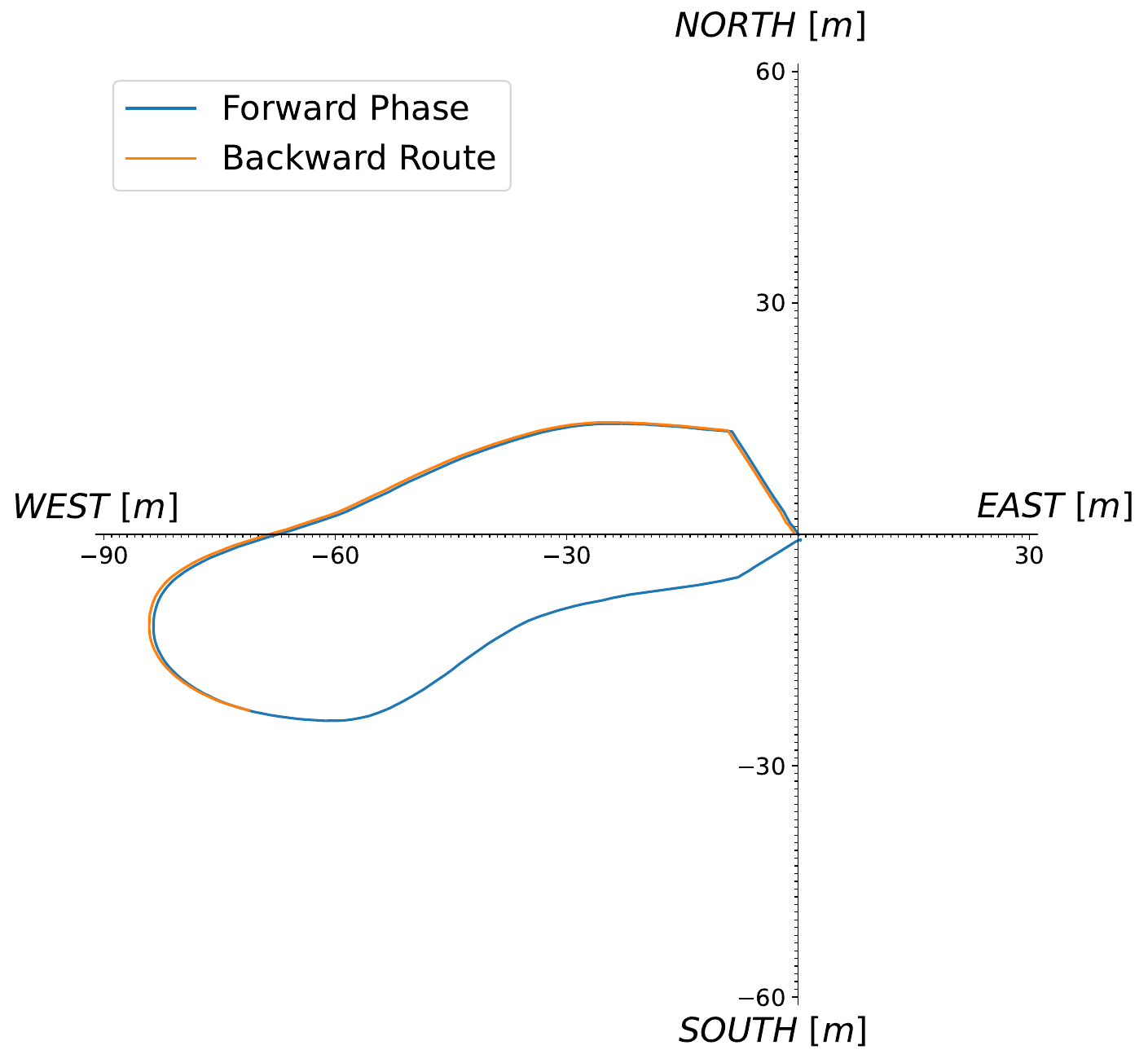}
     \caption{Plot for $\alpha$ = 100\% and $\beta$ = 0\%. In this case $x\_arrive$ = 0m and $y\_arrive$ = 0m.}
     \label{fig:plotAlfaBeta0}
 \end{subfigure}
 \hfill
 \begin{subfigure}{0.24\textwidth}
     \includegraphics[width=\textwidth]{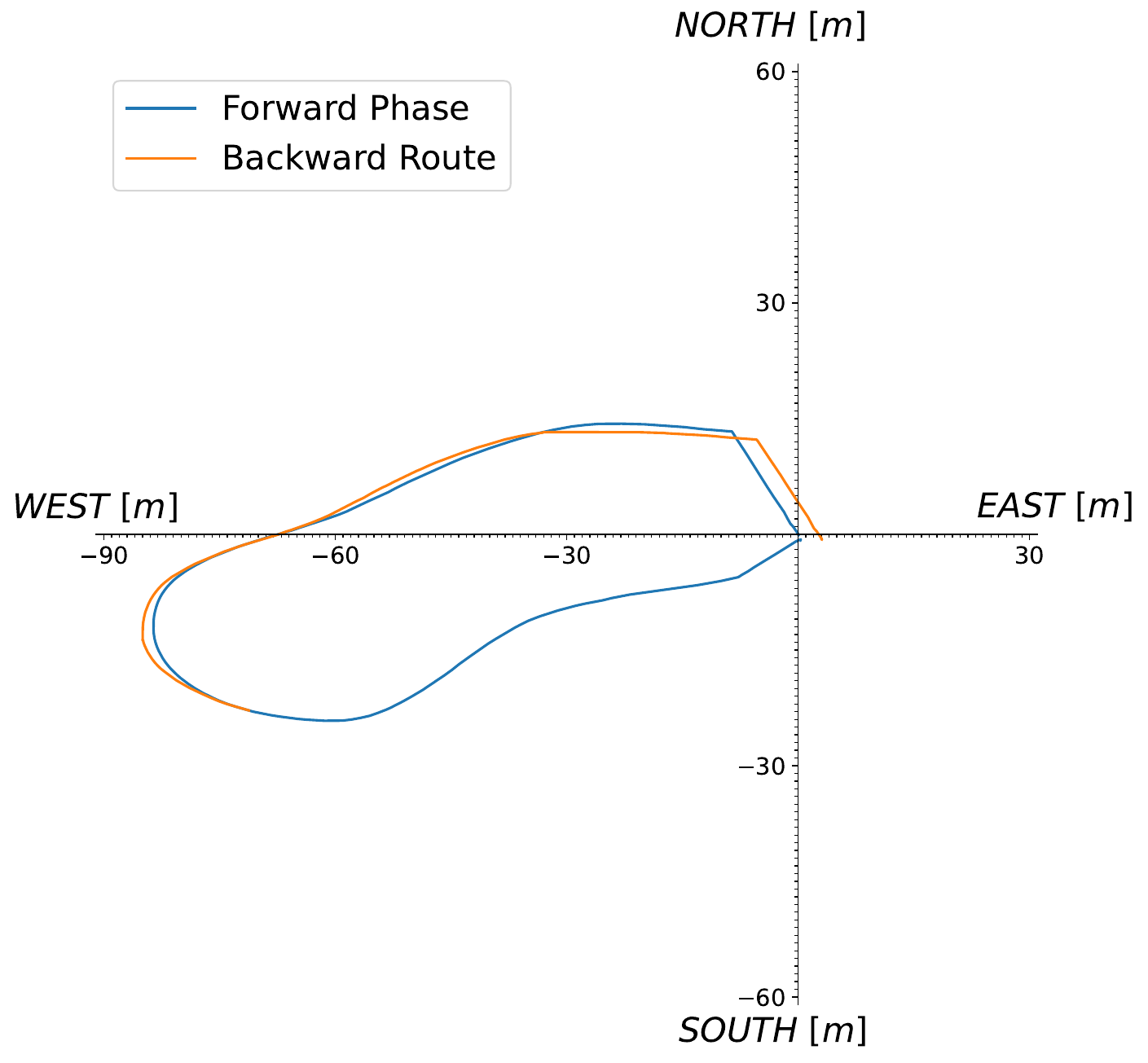}
     \caption{Plot for $\alpha$ = 95\% and $\beta$ = 5\%. In this case $x\_arrive$ = -0.51m and $y\_arrive$ = 2.80m.}
     \label{fig:plotAlfaBeta5}
 \end{subfigure}
 
 \medskip
 \begin{subfigure}{0.24\textwidth}
     \includegraphics[width=\textwidth]{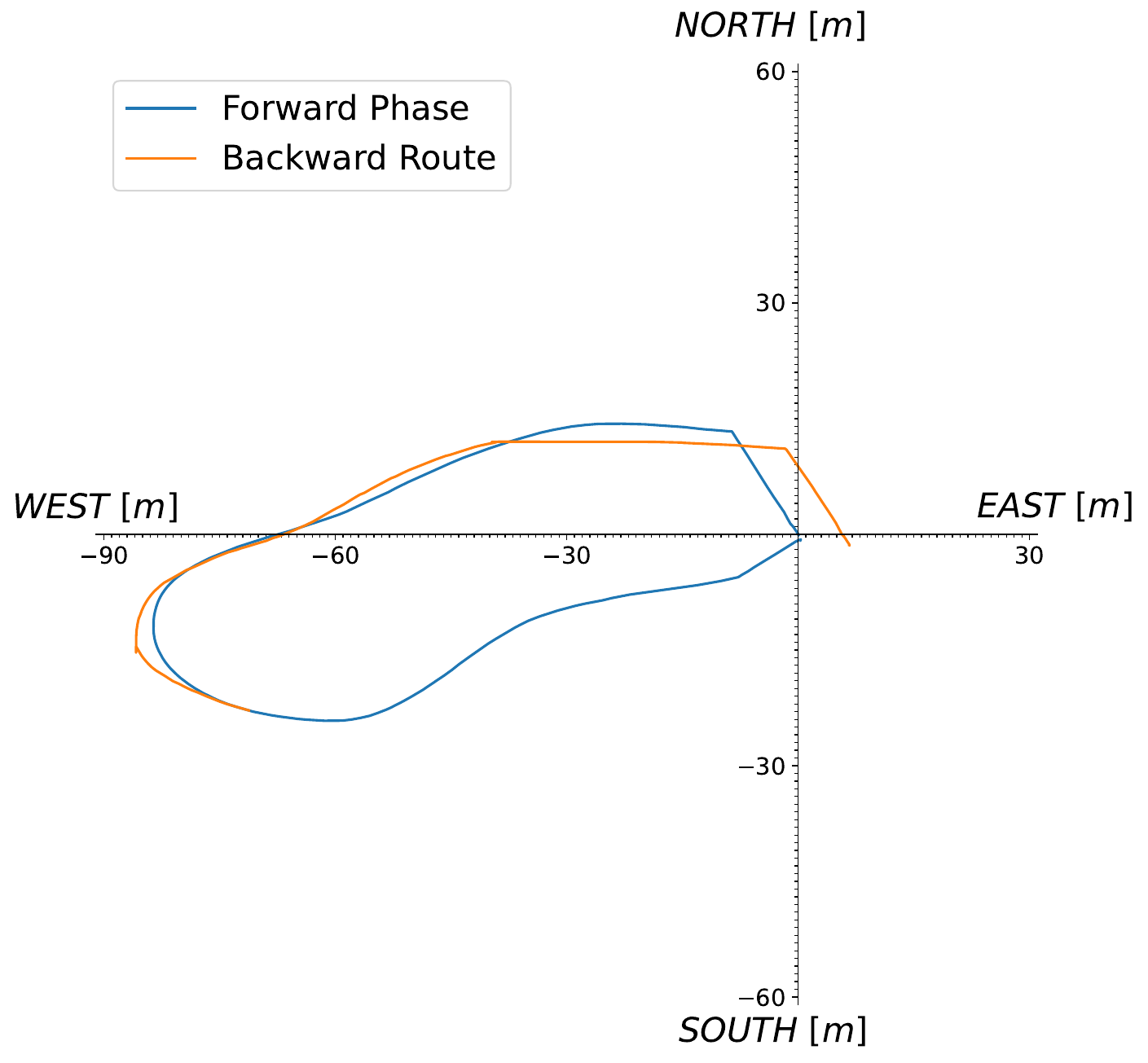}
     \caption{Plot for $\alpha$ = 90\% and $\beta$ = 10\%. In this case $x\_arrive$ = -1.18m and $y\_arrive$ = 6.16m.}
     \label{fig:plotAlfaBeta10}
 \end{subfigure}
 \hfill
 \begin{subfigure}{0.24\textwidth}
     \includegraphics[width=\textwidth]{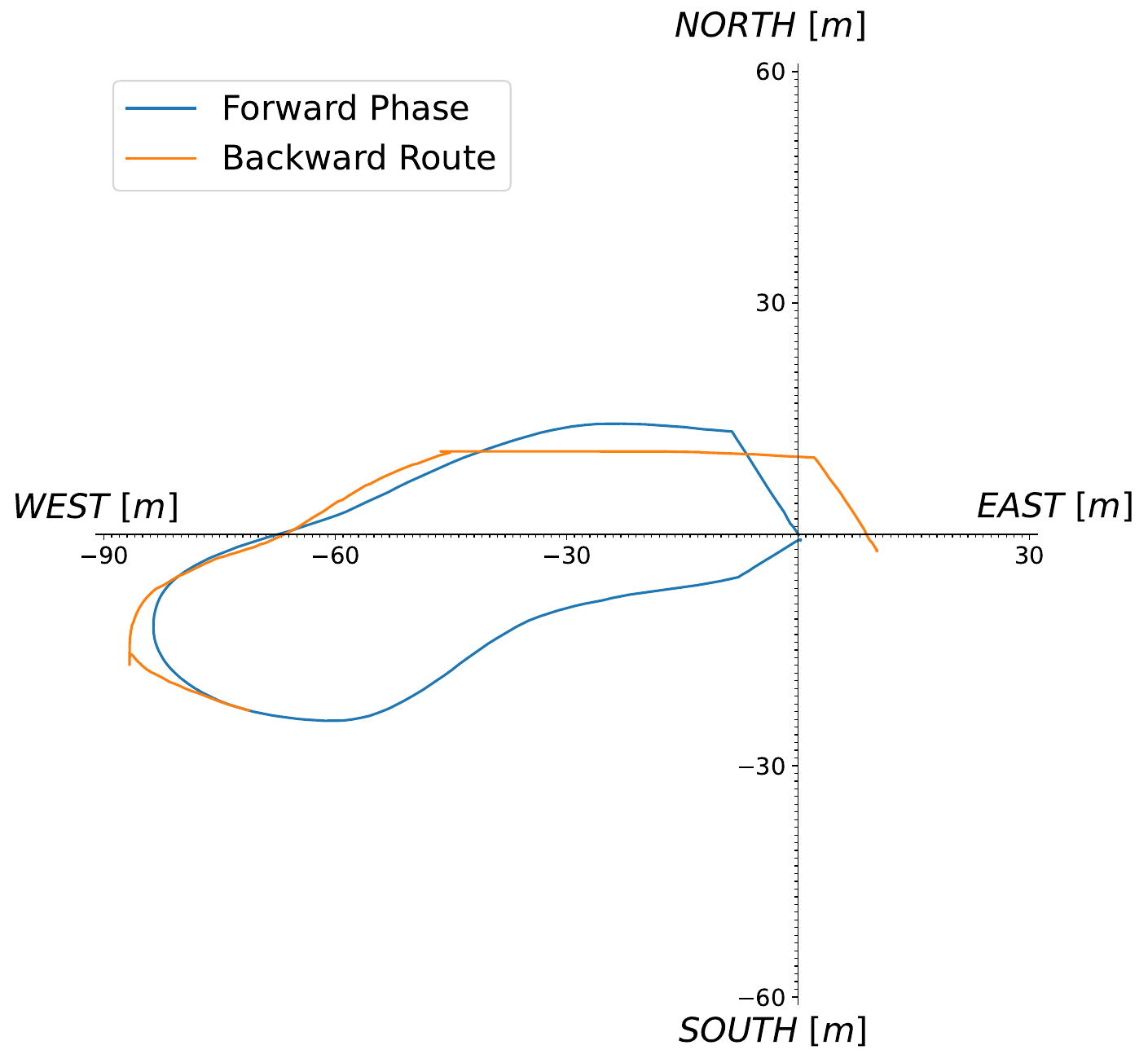}
     \caption{Plot for $\alpha$ = 85\% and $\beta$ = 15\%. In this case $x\_arrive$ = -1.86m and $y\_arrive$ = 9.52m.}
     \label{fig:plotAlfaBeta15}
 \end{subfigure}

 \caption{Backward Phase Evaluation with Weighted Proportion Method ( Section \ref{sec:backward} and algorithm \ref{alg:backpropagation}). Same flight plotted using different value of $\alpha$ and $\beta$.}
 \label{fig:plotAlfaBeta}
\end{figure}

\begin{figure}[t]
 \begin{subfigure}{0.24\textwidth}
     \includegraphics[width=\textwidth]{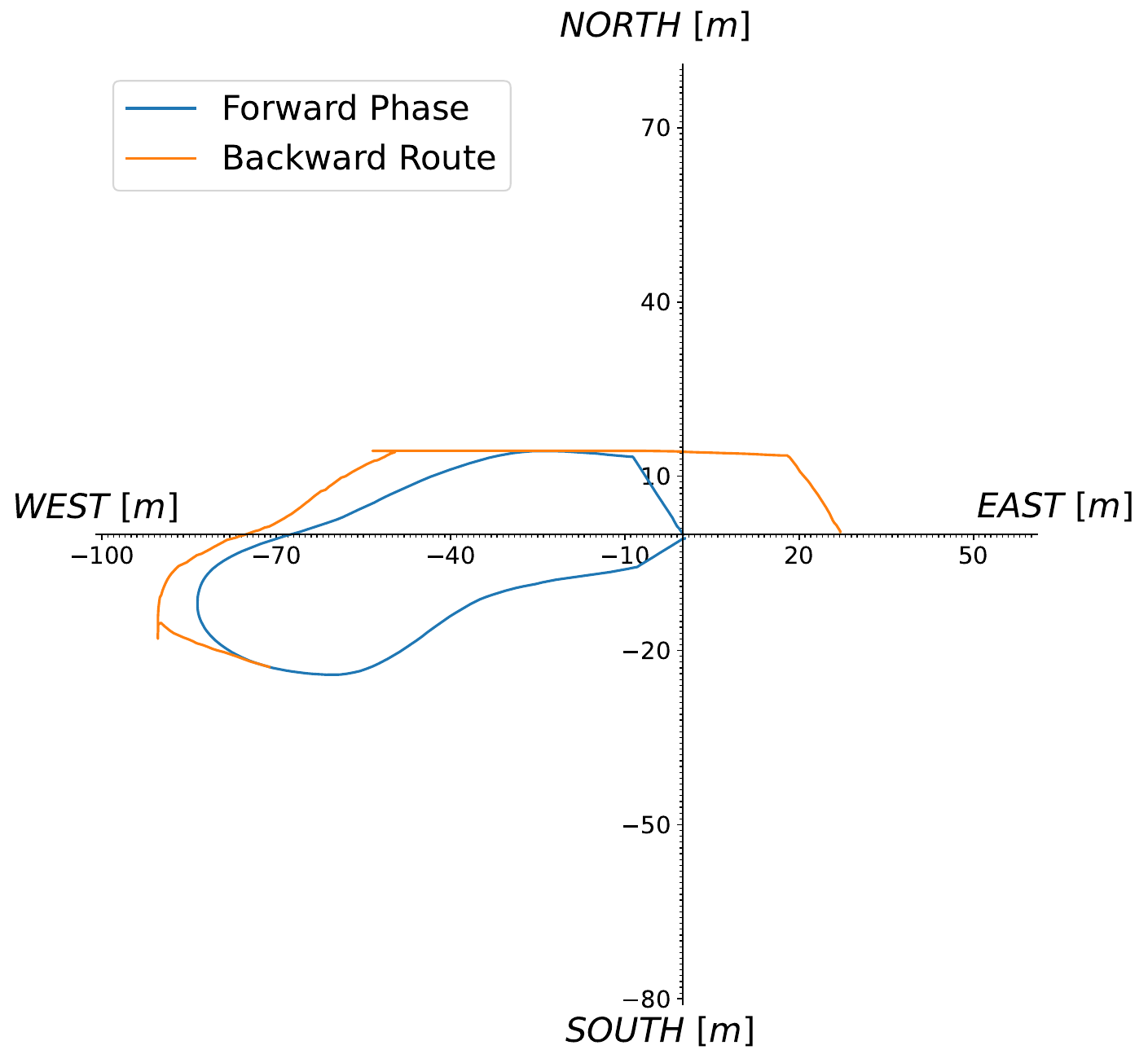}
     \caption{$\gamma$=[2,3]. Each value of the return flight wind is multiplied by a value between 2 and 3.}
     \label{fig:plotWindMultiplied-a}
 \end{subfigure}
 \hfill
 \begin{subfigure}{0.24\textwidth}
     \includegraphics[width=\textwidth]{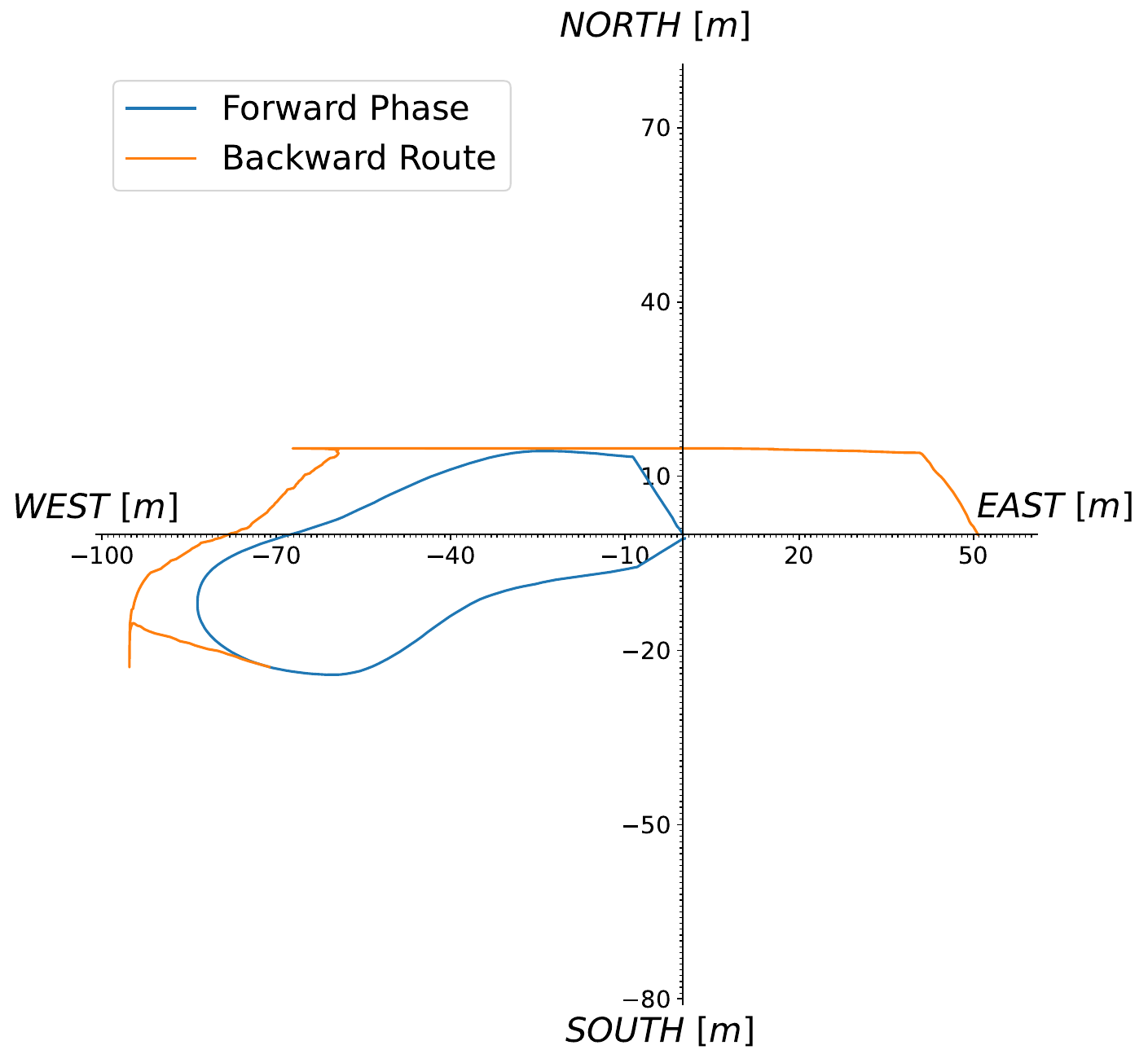}
     \caption{$\gamma$ = [3,5]. Each value of the return flight wind is multiplied by a value between 3 and 5.}
     \label{fig:plotWindMultiplied-b}
 \end{subfigure}
 \caption{Same flight of Figure \ref{fig:plotAlfaBeta10}, with constant $\alpha$ = 90\% and $\beta$ = 10\%, but multiply randomly the wind of the return flight by a factor $\gamma$.}
 \label{fig:plotWindMultiplied}
\end{figure}

\begin{figure}[t]
\centering
{\includegraphics[width=0.3\textwidth]{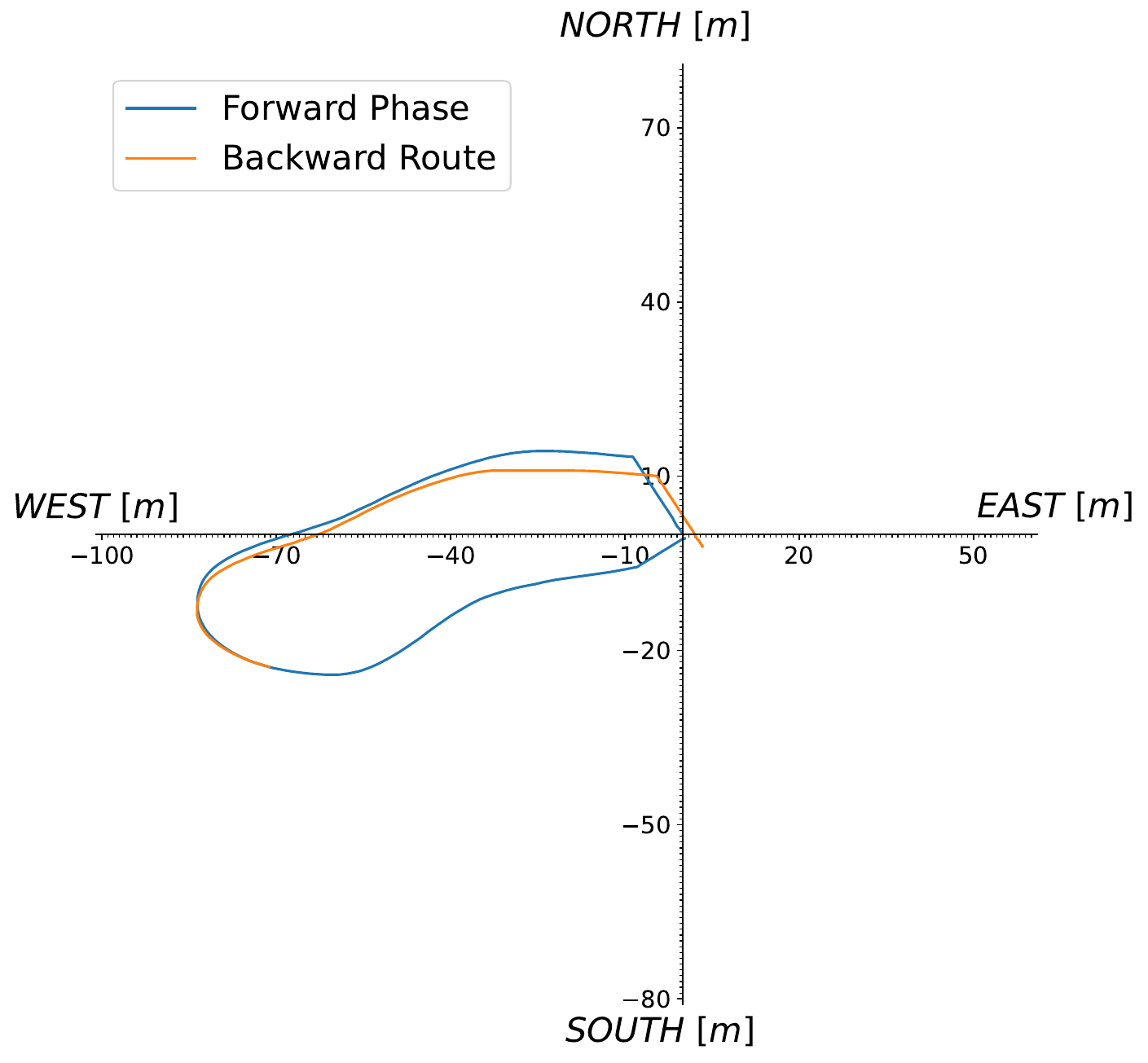}}
\caption{Same path as plot \ref{fig:plotAlfaBeta10} but with the winds of the forward flight multiplied by a factor $\gamma$ = [3,5]. The two graphs are very similar.}
\label{fig:AndataMoltiplicata}
\end{figure}

Starting with the Weighted Proportion Method, in Figure~\ref{fig:plotAlfaBeta} are shown different scenarios where different combinations of weights (i.e., $\alpha$ and $\beta$) are tested. Those pictures show in blue the real path followed by the drone, while in yellow is the solution computed using our system for the backward path. 
For instance, assuming that the GNSS connection is lost (e.g., GNSS spoofing attack or jamming attack taking place) at a certain position, the drone starts the backward phase and tries to redo the forward path in reverse order, that is the yellow line, reaching the starting point.

The ideal case in Figure ~\ref{fig:plotAlfaBeta0}, i.e., no wind presence, therefore, $\beta=0$, shows that the error in the final position is 0 meters both for x\_arrive (i.e., the north direction is the x axe) and y\_arrive (i.e., the east direction is the y axe), meaning that our solution is able to fly back the drone exactly to the takeoff position.
Increasing $\beta$, and therefore the weight associated with the wind impact, shows also increasing error in the final arriving position of the drone, as expected. As stated before, since we think that DJI builds its own algorithm to contrast the wind, we aim to prioritize the xSpeed and ySpeed values over the wind power measured by the anemometer, from our testing we believe that $\beta=10\%$ is the ideal value.
Using this configuration, the average error across all the flights is 14 meters for the x-axis and 3.25 meters for the y-axis, therefore this level of precision is sufficient for the pilot to safely retrieve the drone, minimizing the risk of potential drone theft.

Until now, we only considered a situation where the wind is constant throughout the whole flight, that is the wind speed is the same for the forward path and backward path. In order to test a scenario where the wind speed is disparate, we introduce a random multiplication factor, denoted as $\gamma$, to the
wind data obtained from the anemometer during the forward phase. Figure \ref{fig:plotWindMultiplied} show the results with $\alpha$ and $\beta$ constant, but varying $\gamma$.
As observed, in Figure \ref{fig:plotWindMultiplied-b}, there is a noticeable decline in performance, primarily attributed to the wind intensity during the forward path being approximately 4/5 times less potent than during the return path. It is important to acknowledge that these factors are randomly applied, leading to varying results from one flight to another. To provide a reference for accuracy, Figure \ref{fig:plotWindMultiplied-a} illustrates the drone landing about 28 meters east and 1 meter north of the starting point. In contrast, Figure \ref{fig:plotWindMultiplied-b} depicts a further deterioration in performance, with the drone landing approximately 55 meters away from the starting point. 
It is also crucial to recognize that these assumptions are unlikely to accurately mirror real-world conditions. In practice, it is improbable for the wind's intensity to undergo such drastic changes within the duration of a 10-15 minute flight. Instead, in scenarios where the wind is stronger during the forward route than during the return journey, the algorithm performs effectively as shown in Figure \ref{fig:AndataMoltiplicata}. This effectiveness stems from the fact that, as indicated by eq. \ref{eq:Completa}, the second term tends to become negligible, requiring only the replication of xSpeed and ySpeed values from the forward phase. Also, this seems to back up our hypothesis of DJI using some algorithm internally to counter the wind effect (at least to some degree).

From a computational complexity point of view, on average, the algorithm executes in 5.8 milliseconds. Regrettably, incorporating the algorithm directly into the drone's operations is not feasible. Consequently, evaluating the algorithm in a real-world scenario or determining its execution time is not possible. Nevertheless, it is noteworthy to emphasize that despite the relatively outdated computer used for the algorithm's execution, it consistently achieves favorable execution times, which can be readily considered acceptable for online or on-the-fly solution computations.

\begin{figure}
 \begin{subfigure}[t]{0.24\textwidth}
     \includegraphics[width=0.9\textwidth]{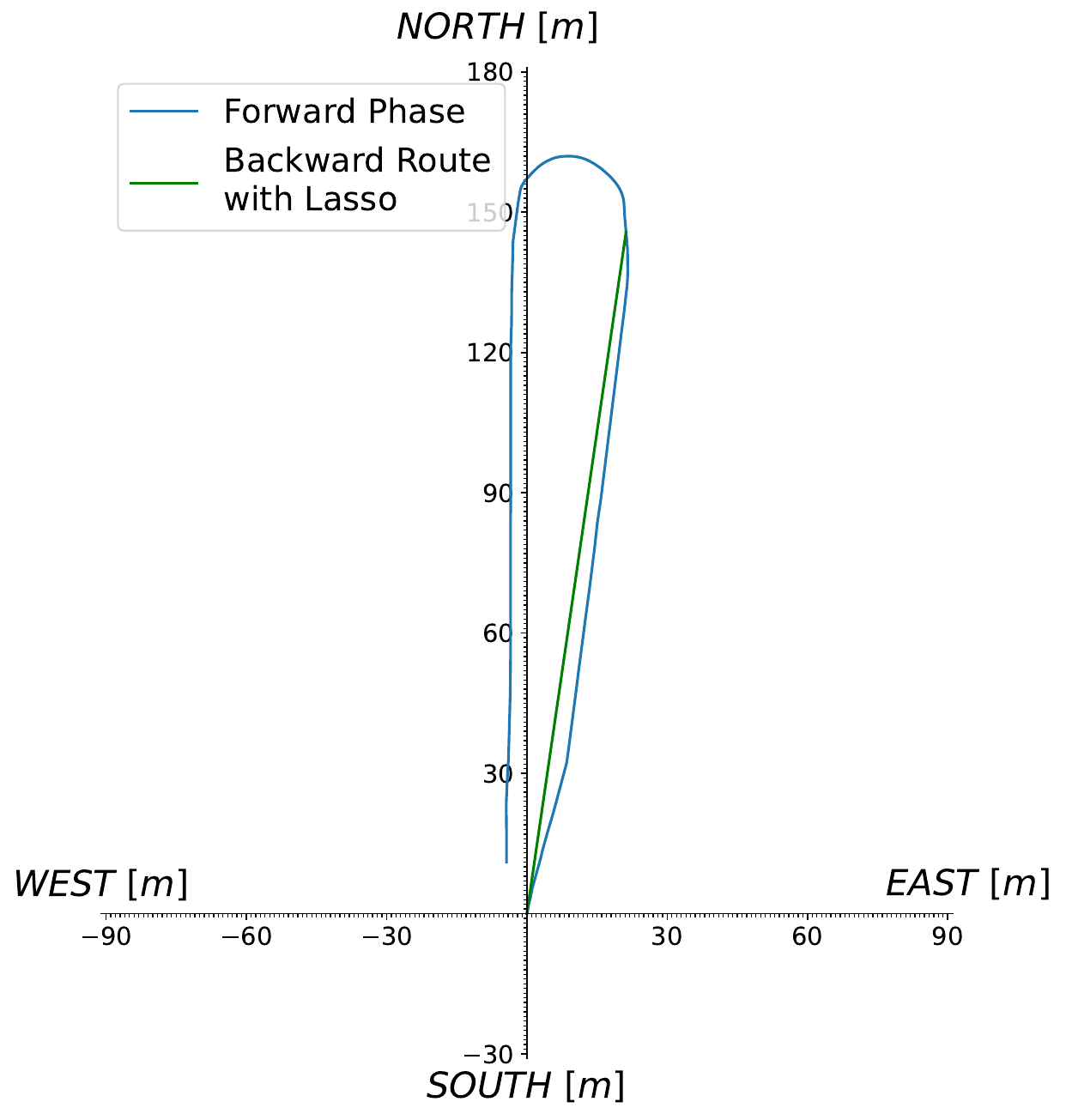}
     \caption{An example of backward route computation using LASSO.}
     \label{fig:lassoback1}
 \end{subfigure}
 \hfill
 \begin{subfigure}[t]{0.24\textwidth}
     \includegraphics[width=\textwidth]{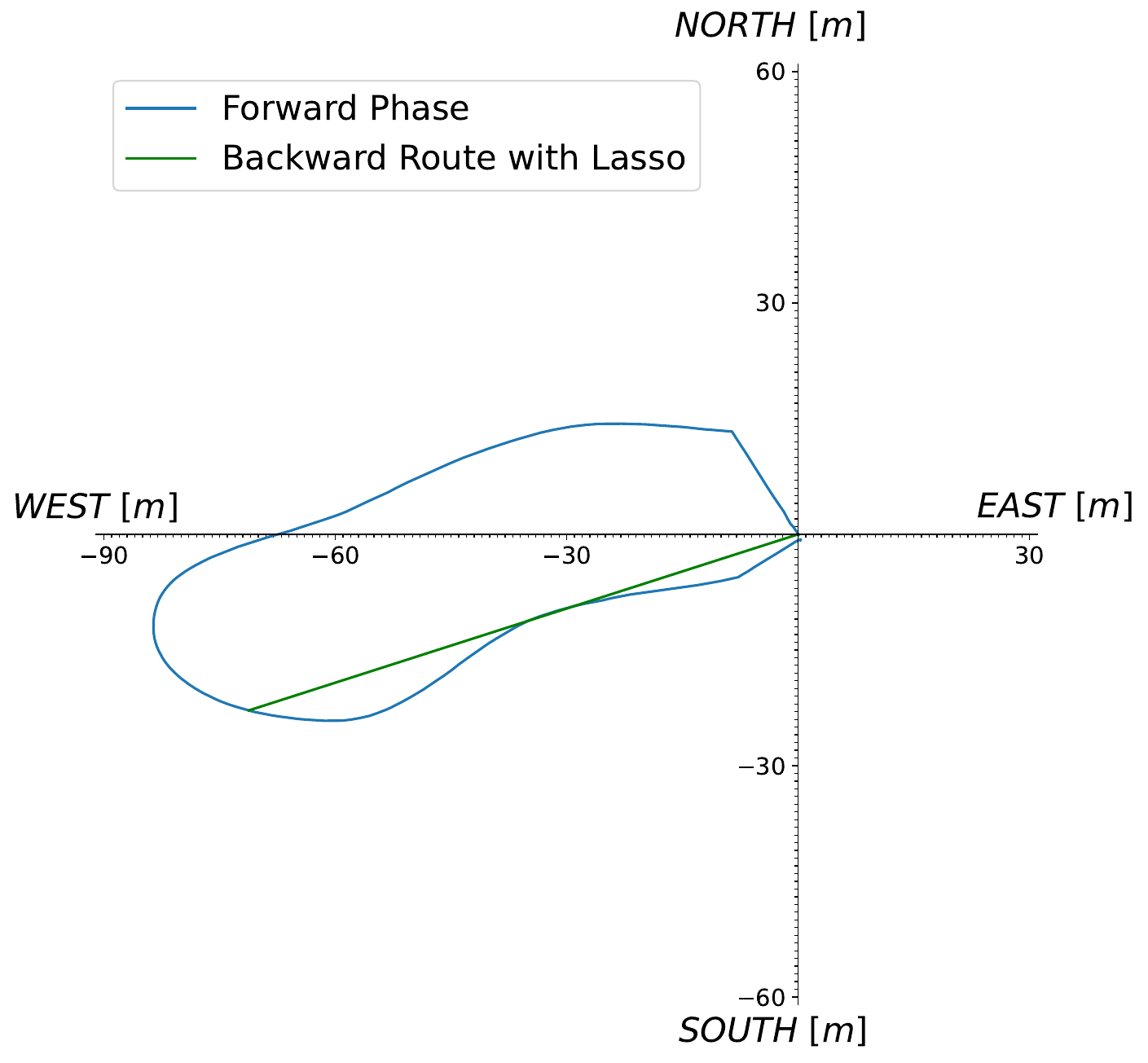}
     \caption{Complex example of LASSO backward route computation.}
     \label{fig:lassoback2}
 \end{subfigure}
 \caption{Representation of LASSO regression on 2 different flights, predicting xSpeed and ySpeed for the backward route.}

 \label{fig:lassoback}
\end{figure}

\paragraph{\textbf{ML-Based Method}}
It is crucial to conduct a proper validation of the predictive model to ascertain its dependability in real-world situations. Figure \ref{fig:lassoback} provides an overview of the algorithm's performance, showing consistent return to the starting point in each experiment. However, it is essential to highlight that this method entails flying in a direct line from the location where the GNSS signal is lost back to the starting point, as opposed to retracing the exact path as the previous method. 

We are aware that this methodology introduces the potential for collisions with obstacles. Even though drones come with an obstacle avoidance system built-in, and that is out of the scope of this work, we have tackled it by programming the drone to automatically elevate its flight altitude, thereby increasing its distance from the ground. As the flight altitude rises, the probability of in-air collisions decreases significantly although the specific risk level can vary depending on the flight environment. Adjusting the flight altitude is a straightforward operation for the drone typically implemented early in the flight. However, for added safety, this elevation can also occur dynamically if the drone is compromised and before starting the Backward Phase.

Suppose the model has been trained offline, and upon completion of model training, it should be integrated into the drone's onboard system. Subsequently, during flights, the model is employed in real-time to estimate xSpeed and ySpeed based on the current northR and eastR values. This approach eliminates the need for retraining the model for each flight, enabling it to offer immediate predictions as required. Under these conditions, the execution time is further reduced compared to the Weighted Proportion Method with an average execution time of approximately 4.8 milliseconds.
It is worth mentioning that an alternative approach is to monitor pairs of [xSpeed, northR] and [ySpeed, eastR] during the entire flight, constructing a distinct model for each flight instance when a potential attack is detected. However, generating a model for every flight may utilize substantial computational resources on the drone, a factor that should be approached with caution. Striking a balance between adapting to changing conditions and ensuring resource efficiency is a crucial point of this work, therefore any consideration regarding the online training of the model is left for future advancements in this field.

\section{Conclusion and Future Works} \label{sec:conslusions}
The absence of a GNSS presents a formidable challenge as it renders the drone unstable, leading to potential crashes. Moreover, a cybersecurity concern is identified, wherein the vulnerability of drones to disruptions in GNSS signals can be exploited through various attacks like Spoofing and jamming. The primary objective of this research is the development of a robust algorithm to enable drones to autonomously navigate back to their initial takeoff location, irrespective of GNSS signals. In this context, wind emerged as a significant factor influencing a drone's flight trajectory. Thus, the Mavic 3 Classic equipped with an anemometer, was acquired for outdoor testing, enabling precise wind-related experiments.

We successfully demonstrated the feasibility of tracking the forward flight path without GNSS assistance. However, accurately retracing the same path during the return journey posed a more challenging task. Two distinct models were simulated and evaluated for this purpose. The first model is based on a relationship where the drone's speed is influenced by the wind, while the second model employs a LASSO method, with the latter proving more effective. However, depending on the particular use case, one solution may be more suitable than the other: for instance, if there's an opportunity to gather data in advance (like patrolling an area), the ML-based method should be chosen, whereas the Weighted Proportion Method is more suitable for a one-shot scenario. Furthermore, these two proposed methods can be used together; for example, starting with the first method during the initial flight to collect data and then switching to the second method afterward.
Unfortunately, real-world scenario evaluations are complicated due to restrictions on installing external software imposed by DJI, potentially leading to legal and insurance complications.

In essence, our proposed system achieves GNSS independence not because the features we employ are inherently GNSS-independent, but rather because following the compromise of the drone through a GNSS attack, it can autonomously recover by leveraging the path history up to the point of the attack. Consequently, the solution becomes entirely disengaged from GNSS signals once the attack is successful.

As main potential future researches, we envision use cases where there are extreme Wind Conditions. Evaluating the algorithm's performance in challenging windy environments, such as atop a mountain with strong gusts, could yield valuable insights. However, due to the homemade nature of the setup, caution is warranted to avoid potential crashes.
Another very important aspect is the in-depth analysis of Rotor Data. In fact, examining the behavior of rotors and their correlation with xSpeed, ySpeed, and wind power could be a promising topic for future research. However, accessing proprietary rotor data may pose challenges due to its sensitive nature. Probably the easiest way to improve the proposed system is to employ more advanced ML models, like neural networks and deep learning but those require an extensive data collection phase, also considering different wind situations.

Finally, an inboard implementation would be desirable. Flashing the algorithm inside the drone to test it in a real scenario represents both a significant opportunity and a formidable challenge. This step is crucial for practical application and real-world testing, especially from a computational and execution speed point of view.
Exploring these directions for future research can deepen our understanding of how algorithms adapt to diverse scenarios and enhance their overall performance.

\bibliographystyle{IEEEtran}
\bibliography{bib}

\end{document}